# AI and the Opportunity for Shared Prosperity:
# Lessons From the History of Technology and the Economy

Guy Ben-Ishai, Jeff Dean, James Manyika, Ruth Porat, Hal Varian, Kent Walker[+]

January 31, 2024

## Abstract

Recent progress in artificial intelligence (AI) marks a pivotal moment in human history. It presents the opportunity for machines to learn, adapt, and perform tasks that have the potential to assist people, from everyday activities to their most creative and ambitious projects. It also has the potential to help businesses and organizations harness knowledge, increase productivity, innovate, transform, and power shared prosperity. This tremendous potential raises two fundamental questions: (1) Will AI actually advance national and global economic transformation and growth to benefit society at large? and (2) What issues must we get right to fully realize AI's economic value, expand prosperity and improve lives everywhere? We explore these questions by considering the recent history of technology and innovation as a guide for the likely impact of AI and what we must do to realize its economic potential to benefit society. While we do not presume the future will be entirely like that past – for reasons we will discuss – we do believe prior experience with technological change offers many useful lessons. We conclude that while progress in AI presents a historic opportunity to advance our economic prosperity and future wellbeing, its economic benefits will not come automatically and that AI risks exacerbating existing economic challenges unless we collectively and purposefully act to enable its potential and address its challenges. We suggest a collective policy agenda – involving developers, deployers and users of AI, infrastructure providers, policymakers, and those involved in workforce training – that may help both realize and harness AI's economic potential and address its risks to our shared prosperity.

* Equal contributions, the authors are affiliated with Google and Alphabet.

Table of Contents

## INTRODUCTION

AI is not merely about chatbots – it is much broader and has applications from the everyday to the extraordinary. It is a foundational technology that is increasingly powering and improving many tools, applications, products and services – including those not normally thought of as AI – and enabling new ones with new powerful and useful capabilities. In this way it is a foundational technology that has the potential to assist people to be more productive and creative, power small and large businesses, transform sectors, drive overall productivity, boost economic growth, and thus change the trajectory of national and global economies. It also has the potential to help advance science, help address some of society's most pressing challenges, and contribute to our overall prosperity and wellbeing.

However, the creation, implementation, and scaled use of technology isn't like flipping a light switch; it's not a single moment of discovery. It is a long, fragile, and often complex process. It takes decades of combinatorial innovation, policy choices, and sustained efforts to simultaneously enable its potential and address its complexities and risks. And its success is never really guaranteed. Like other technological advances, AI's journey from the lab to real-world usefulness has been lengthy and is still fraught with a multitude of obstacles. Moreover, it is unlikely that humanity will realize AI's full economic potential unless we invest in infrastructure, commit to an innovation ecosystem, diffuse its use across the economy in productivity-enhancing ways, help our workforce transition, empower our small businesses, develop trustworthy products and services, and enable all to participate in its beneficial possibilities.

Will this profound new technology actually lead to technological inflection points, productivity growth, and opportunities to advance our economic prosperity and future wellbeing? To address this question, we draw upon the history of technology and innovation and examine its potential to accomplish such goals. We also assess the possibility that while AI has the potential to improve many economic aspects such as growth, lowering barriers to knowledge and expanding opportunity, AI could also cause adverse economic effects, including widening income inequality and displacing workers, as both the reality and perception of these risks could slow or block AI's wide adoption and economic benefits.

While there is a lot that we must collectively get right about AI, including issues of safety and responsible AI development, in this paper, we focus on the economy and shared prosperity and explore four key questions: (a) Why AI's economic potential to improve our prosperity is immense, (b) Why realization of its economic potential and benefits is not guaranteed, (c) What history tells us about the relationship between technology and jobs, and (d) What collective policy action is needed to realize AI's economic potential and address its challenges.

1. WHY AI'S ECONOMIC POTENTIAL IS IMMENSE

We start by posing a sweeping but seemingly simple question: Can AI change the global economic landscape and expand the frontier of economic opportunity? We'll argue that AI marks a clear departure from prior technologies, potentially meeting the criteria to become a general-purpose technology. We then consider how to take advantage of AI's unique qualities to advance key mechanisms of economic progress, such as improving labor productivity, transforming work and expanding economic opportunity, helping advance scientific discovery, and tackling pressing societal challenges.

**Attributes of general-purpose technologies**

Throughout modern history, there have been only a handful of technologies, such as the steam engine, electricity, and personal computing, that economists have labeled as general-purpose technologies (see Bresnahan and Trachtenberg). These have been applied across a wide range of activities by individuals and organizations spanning across all parts of the economy and society in productivity-enhancing ways. What enables a single technology to have such a transformative economic impact? There are three distinct features that are commonly identified as the building blocks of general-purpose technologies: (a) broad applicability, (b) large innovation spillovers, and (c) technological improvement. AI appears to have the potential for all three.

Broad applicability: A defining attribute of general-purpose technologies is their pervasive application across a spectrum of economic activities, ranging from those undertaken by individuals to those carried out by small and medium-sized businesses (SMBs) to large companies and organizations across all industries and areas of economic activity. Similar to electricity and the steam engine, AI is a prime candidate for widespread use because it offers nearly universal capabilities and applicability. As a tool for recognizing patterns and making predictions (see Agrawal et al), AI can be used to improve the performance of many tools, systems, products and services, support an endless range of decisions, solve complex problems, and help in the design and development of applications in a wide range of environments. And its capacity to enable machine perception (through processing natural language, recognizing speech, and understanding images) can dramatically expand the environments where it can be successfully applied.

Over the last couple of decades machine learning, a core element of modern AI, has become increasingly pervasive – often in ways that most users would not think of as AI systems, including in smart-phone features, search, natural language processing, language translation, image analysis, biomedical equipment, fraud detection, and quality assurance. As AI's power and capability has advanced, especially over the last decade (see Dean), its potential for wider application and use across business functions has grown, in both routine tasks and knowledge work.

Recent studies by many researchers map AI's potential across business functions and corroborates the potential for its wide-ranging applicability, with AI in one study creating significant value over existing analytical tools in 69% of business use-cases and emerging as the sole analytical solution in an additional 15% (see Chui et al). Consistent with these findings, numerous studies find that AI has been successfully applied across dramatically different business functions including mitigating code complexity (see Tabachnyk et al) and enhancing software development (see Kalliamvakou), improving customer support through conversational guidance (see Brynjolfsson), generating sales leads (see Nan et al), improving financial forecasts of stock (see Cao et al), reviewing radiology and medical imaging (see Agarwal et al), and optimizing taxi dispatch by locating real-time, customer-saturated routes (see Kanazawa et al).

Similarly, empirical research into work tasks and occupational exposure also indicates that AI can be applied to tasks in the vast majority of occupations and across nearly every economic segment. For instance, a recent study that aggregated survey data across all U.S. occupations found that one-fourth of work tasks could be automated by AI, with particularly significant opportunities in administrative (46%) and legal (44%) professions (see Briggs and Kodnani). A similar study concluded that AI can automate tasks associated with 30% of hours worked across the U.S. economy (see Ellingrud et al). And additional studies (see OECD and Eloundou et al) also found significant and increasingly general applicability of AI across tasks in nearly all occupation groups and economic segments. Moreover, such studies not only corroborate AI potential for broad application across diverse occupations and business functions, but also indicate that such application can lead to significant economic impacts. Several academic, think tank and business research studies suggest significant economic potential from AI via various channels of impact (see European Parliament Research Service). For example, although they do not explicitly state how long it would take, McKinsey Global Institute estimates that AI (including generative AI) could deliver $17 to $25 trillion annually to the global economy (see Chui et al). Goldman Sachs projects that AI could increase global  GDP by 7% (see Briggs and Kodnani). But such estimated gains are not automatic as we will discuss later.

Innovation spillovers: Another important attribute of general-purpose technologies is that they lead to a cascade of complementary innovation and technologies spillovers. Contrary to other technologies that may only trigger new innovations once at a discrete point in time, general-purpose technologies open up new opportunities for continued innovations. These opportunities often take place due to a positive feedback loop between the development of applications and advancement of associated science and technologies (see Bresnahan). For example, the advent of PCs gave rise to a continual stream of new innovations including digital navigation, e-commerce, and social media, while electricity led to a plethora of related innovations, upstream advances in power generation and grid development and downstream advances in existing consumer products like the telephone, TV, and household appliances.

Because AI product development depends on advances in science and technology and has the potential to transform so many aspects of human endeavor, its development can serve as a catalyst for unlocking further frontiers of innovation. This potential is reflected by the flood of VC investments, scientific collaborations, and thriving AI ecosystem of researchers, developers, startups, infrastructure, tools and services already triggered by AI in recent years (see Stanford AI index). In the same way the printing press led to the Scientific Revolution, the steam engine led to the Industrial Revolution, and the transistor led to the Digital Revolution, AI can lead to combinatorial innovations and compounding benefits that we may not fully envision today.

Technological improvement: The third commonly-cited attribute of general-purpose technologies is the possibility of continual improvements in their performance and capabilities over time. Various research projects have been tracking the continual and rapid advances of AI (see Stanford AI index). These improvements come from innovations born out of prior technological achievements and accumulated scientific knowledge. Many are also the result of scale (see Bloom et al), where R&D intensity improves scientific success, as well as network effects, where the value of each innovation increases as more people or organizations use it, such as the value of a fax machine or a telephone (see Hagiu and Yoffie, Shapiro and Varian). Personal computers exemplify how continual and compounding improvements work: Where, over decades, the power of computing has continually doubled roughly every 18 months (since Robert Moore made this observation in 1965), cars and trains have not experienced the same rapid exponential improvements sustained over decades (e.g., in horsepower, fuel efficiency or haul).

In recent years, as machine learning, statistical techniques, compute power, and large data availability have improved (see Gelman, Our World in Data), so has the performance of AI models in dimensions that are both anticipated and unanticipated (see Wei et al), and while there is some debate as to whether these are truly emergent or not (see Schaeffer et al), the scaling and expansion of capabilities is clearly evident. As new model innovations are developed, unit compute costs decline, more compute is applied to model training, and scaling laws (see Kaplan et al) continue to hold, we can expect the performance and capabilities of the transformer-based AI systems to continue to improve. Moreover, growing usage of early AI systems is also making it possible to learn and improve from user feedback and techniques such as reinforcement learning from human feedback (see Steinnon et al). And similar to prior technologies that benefited from network externalities, services powered by AI models improve these models through the use of implicit or explicit user feedback data, thereby attracting new users and generating a similar virtuous cycle (see Levine and Jain, Clough and Wu).

To clarify the analysis above, we recognize that the threshold for becoming a general-purpose technology is understandably high. And although there have been plenty of candidates over the years (e.g., blockchain, nanotechnology and the internet of things), many of these have so far proved to be limited in their generality. Only a handful of technologies have actually been broadly applied and adopted across business activity and across sectors, and have generated the productivity-enhancing, economy-wide benefits required to meet the definition. Yet the foundational nature and paradigm shift that AI presents, along with its clear potential for widespread application and expansion of the innovation frontier, leave us optimistic about its tremendous potential to generate benefits at a general-purpose technology level. It appears that a number of other studies (see [Chui et al](#), [Brynjloffson et al](#), [Goldfarb et al](#), and [Crafts](#)) have reached similar conclusions about AI's potential as a general-purpose tool.

**A clear departure from prior digital technologies**

<u>What features does AI share with other digital technologies?</u> Digital technologies have already established a high bar for widespread transformative impact, illustrating how critically important information, knowledge and communications are for economic development. For instance, in the late 1990s, ICT technologies played a key role in driving widespread productivity growth that helped resolve the so-called Solow Paradox (see [Farrell et al](#), [Triplett](#)). And while their impact on productivity slowed in the decade before COVID (see [Remes et al](#)), the potential re-emerged during the COVID era (see [Manyika and Spence](#)), as companies embraced new digital technologies. Beyond the wide productivity gains, over the past decade and a half, the digital economy outpaced the rest of the economy by a factor of 3.6 (see [Bureau of Economic Analysis](#)). Moreover, digital technologies not only reduced the costs of accumulating knowledge, they more importantly fueled new forms of innovation such as e-commerce, cloud computing and social media (see [Brynjolfsson and McAfee](#)). Digital technologies have been impactful because of several distinct economic features: scalability, widespread applicability, low replication costs and low marginal costs, as well as their, in principle, non-rivalry characteristics (see: [Jones and Tonetti](#)). AI shares and builds on these features.

<u>What AI features are new?</u> Unlike previous digital and automation technologies that require precise, step-by-step instructions to perform tasks, modern AI can learn from examples and patterns and recently in self-supervised ways, to perform actions even when they are not explicitly coded or prompted. This has led to a number of important benefits, ranging from improved pattern recognition and enhanced predictions to expanded automation of non-routine and cognitive tasks (see [Brynjolfsson and McAfee](#), [Agrawal et al](#)). Another set of new innovations in AI are Large Language Models (LLMs), systems that are capable of tackling an increasingly wide range of tasks across a diversity of topical domains and modalities such as text, software code, audio, image and video. These expanding capabilities have led some to describe LLMs as foundation models (see [Bommasani et al](#)) now approaching wide levels of generality (see [Aguera y Arcas  and Norvig](#)).

Modern AI can also be distinguished from prior technologies by its increasing ability to perform at near-human expert levels across many cognitive tasks and benchmarks (see

Hendrycks et al), including CFA exams (see Callanan et al), disease diagnosis (see Rodman et al), and even standardized creativity tests (see Guzik). And AI's growing ability to use other software and systems allows it to serve as a cross-connector across existing digital and physical applications. AI has also improved robotics through integration with transformer-based LLMs (see Driess et al, Brohan et al), progress in robotics however (and other systems operating in the physical world and especially in unstructured environments) has not been as rapid as that in knowledge-based tasks. This differential rate of progress is important to keep in mind when considering impacts on the economy given that a significant component of work in the economy involves activity in the physical world.

While improvements in AI have been significant, current AI systems, especially LLM-based systems (see Weidinger et al, Lappin), have some performance and other limitations with respect to factuality and hallucinations, and in some cases safety, or capability gaps with respect to memory and planning that affect their broad applicability. These however are all areas of ongoing research.

**Mechanisms of economic progress**

Given that AI represents a clear shift from prior technologies and shares many of the attributes of general-purpose technologies, we now turn to the question of how it may advance mechanisms of economic progress. Here we consider several key possibilities: (a) Improving productivity and economic growth, (b) Transforming work and expanding economic opportunity, (c) Advancing scientific discovery, and (d) Helping address pressing societal challenges that could threaten collective prosperity.

**(a) Improving productivity and economic growth**

Labor productivity, a central concept in economics, is a key determinant of a country's standard of living. It is usually defined as output per labor hour and can grow by improving the numerator (i.e. the value of economic output) and/or lowering the denominator (i.e., the total hours worked). Improvements in productivity can occur at the level of a company, sector, or the whole economy, and are most impactful when they occur across many companies and sectors of the economy, especially the large ones. Productivity growth is crucial for increasing wages and living standards, helping raise the purchasing power of consumers, and increasing demand for goods and services. By leading to more and better-paid jobs, and expanding employment possibilities and career paths, productivity gains can also contribute to workers' overall economic security.

Productivity's contribution to overall economic growth is critical and will become more so going forward. Gross Domestic Product growth is largely the product of growth in productivity and growth in the labor supply. Therefore as societies age and/or their population growth slows, the importance of productivity growth for economic prosperity becomes crucial. And when productivity growth is numerator driven, meaning that the value-added output increases faster than reduction in labor inputs (as opposed to simply reducing labor inputs without an increase in value-added output) (see Manyika and Spence), it creates a virtuous cycle of employment growth to meet demand and growth in the value of the output from these sectors, growth of the economy and overall prosperity.

While labor productivity serves as the bedrock of long-term sustainable economic growth, it has experienced two dramatic setbacks in recent decades (see Manyika and Spence). First, several advanced economies have faced long-term declines in productivity growth. In the U.S. productivity growth, which averaged 1.7% in the decade leading to the financial crisis, dropped to 0.5% in the decade before the pandemic (see Remes et al, Manyika and Spence, and Bureau of Labor Statistics). Second, labor productivity growth has been especially sluggish in specific sectors, some of which make up a large portion of the economy. For example, large service sectors — which account for nearly 80% of U.S. employment — achieved pre-pandemic productivity growth of just 0.16%. Such sectors not only witnessed dismal productivity gains, they are also thought to have contributed to innovation bottlenecks and hindered economy-wide progress (see Acemoglu et al). For these reasons, a technology that facilitates productivity-enhancing opportunities across multiple sectors, especially the large ones, holds an enormous potential to restore overall economic growth.

Thus the fact that productivity growth has been sluggish in most advanced economies should be of concern. And the potential of AI to significantly contribute to improving productivity becomes of paramount importance, especially as productivity will now have to do more "work" in the absence of expansions in the labor supply. This is because as populations age and their labor supply declines, their ratio of dependents (i.e., people younger than 15 or older than 64) to the working-age population (i.e., those that are between 15 and 64) tends to grow.  In fact, over the past decade aging economies such as China (18%), the U.S.(9%) and the OECD (8%) have experienced this phenomenon. Thus, absent a productivity boost, these demographic changes are expected to drag economic growth throughout the OECD and in China (see World Bank).

Recent case studies on AI and productivity (e.g. Baily et al, Yee et al) are cause for some optimism. They suggest that AI is poised to help increase productivity across a wide range of occupations (e.g. developers, call centers employees, telemarketers, financial analysts and radiologists), and business functions (e.g. sales, logistics, and operations). The case studies suggest that such gains are not only potentially prevalent across many parts of the economy but also quite large. For instance, in some studies, call centers reported an improvement of 14% in issues resolved per hour (see Brynjolfsson et al), developers completed assigned tasks 55% faster (see Kallimavakou), and AI-assisted financial analysts outperformed other human analysts in 57% of the forecasts (see Cao et al). In another study researchers found that businesses that used relatively basic versions of AI grew revenues 40% faster and employment 25% faster than those that did not (see Alderucci et al). The same study also found that businesses using AI witnessed accelerated growth in value-added per employee. Such early results are encouraging.

Additionally, a number of broad studies indicate that the gains observed in the case studies (which focus on individual use cases or experiments), may be generalizable to the rest of the economy (see Baily et al). One study by McKinsey Global Institute estimated that generative AI could increase U.S. labor productivity by 0.5% to 0.9% a year (relative to an average annual growth of 1.4%) and concluded that productivity gains are the biggest potential economic benefit of generative AI (see Chui et al). Another study released by Goldman Sachs estimates that generative AI alone could boost U.S. labor productivity by 1.5 percentage points a year, roughly the same size increase that followed the emergence of prior transformative technologies like the electric motor and personal computer (see Briggs and Kodnani). Such findings have led to estimates that AI could potentially double overall labor productivity (see Colvin). However, many have been quick to point out that these gains will not happen automatically, a topic we will discuss in Chapter 2 of this paper.

### (b) Transforming work and expanding economic opportunity

AI also has the potential to transform the very way we work. By automating tasks that tend to be more mundane or unsafe, AI has enabled humans to enjoy autonomy at work and focus on endeavors that require more expertise, judgment, creativity, and intuition. In fact, we have already witnessed a similar dynamic prior to AI. Over the last six decades, during a period of increased automation, the share of occupations that entail critical thinking, problem-solving and adaptability continually grew: up from 6% in 1960 to 34% more recently (see Deming). These are important dimensions of work quality (see OECD Employment Outlook). Perhaps one of the most interesting findings from early adoptions of AI is that it can actually lead to a measured increase in both productivity and job satisfaction, often derived from the elimination of mundane or repetitive tasks or the capacity to pursue more knowledge-based tasks. For instance, a recent OECD study (see Lane et al) found that the majority of employees using AI at work reported that in addition to being more productive, AI had also improved their enjoyment of work (63%) and overall mental health (54%).

AI can also play an important role in advancing economic opportunity by making technology accessible to a wider range of users who can utilize it in productive and creative activities – thus making AI a powerful engine for broad economic opportunity. Furthermore, the ability for users to interact with AI in ordinary language, often requiring little initial expertise by the user, can effectively lead to a more inclusive economy by simultaneously improving its utility and broadening access to digital technologies and to the world's knowledge. There are a number of areas where AI can play such a role, like expanding access to skills, empowering small businesses, and turbo-charging progress of many in emerging markets as we now discuss:

<u>AI-assisted work and workforces</u>: AI can play a significant role in democratizing access to relevant work skills and occupations. As most practical uses of AI are at the level of tasks (e.g. Acemoglu and Restrepo, D'Andrea Tyson and Zysman), it presents opportunities to assist workers in a wide range of occupations. Generative AI tools in particular can bridge the knowledge gap by acquiring information accumulated by experienced and expert employees and disseminating it to less skilled or experienced employees (see Agrawal et al). That widens access to expertise (e.g., coding, creative writing, or language skills) that were often the exclusive provenance of workers with more experience or better education (see Brynjolfsson et al, Eastwood). Generative AI tools can also expand access to occupations when used as a collaboration tool for less experienced or technically-trained workers. For instance, while many developers use AI collaboration tools for code completion and bug detection, less experienced developers often report disproportionate benefits from such tools (see Dohmke). These capabilities are particularly important in scenarios where formal education, legacy preferences, or training gaps serve as economic mobility barriers.

<u>Small and medium-sized businesses</u>: Over the last two decades, digital technologies have played a critical role in the success of small and medium-sized businesses. Through the use of digital tools, SMBs gained access to critical business functions that for decades have been either prohibitively expensive or infeasible at small scale. For instance, the ability to run a marketing campaign, the exclusive domain of large firms that could afford an advertising agency in the 1960s, became broadly available through digital marketing tools. And the ability to export products, which historically required elaborate shipping, tariffs, and logistics functions, are now available through basic e-commerce tools. By broadening access to such functions, digital tools had a profound impact on the ability of SMBs to compete more effectively, operate more efficiently, and access markets that span far beyond their immediate geographic proximity. Moreover, leveraging digital technologies to empower SMBs can result in wider economic impact. In fact, several studies have shown that boosting the productivity of SMBs across the economy is important to driving overall productivity (see Gu, OECD).

Because of its wide-ranging applicability and use across a multitude of business functions, AI can dramatically expand such digital impacts as well as impacts from its additive capabilities (see Chui et al). By leveraging advanced pattern recognition, predictions, natural language and image processing, AI is poised to advance access to numerous new applications and business functions such as knowledge management, personalized training, inventory analysis, or even product development. Already today, startups and entrepreneurs are leveraging  access to such functions through cutting-edge AI applications that can be easily customized, scaled and integrated. This potential is fueling growth in funding not only for startups focused on AI's continued development, but also for AI-enabled startups focusing on applying it to a wide variety of domains and use cases ranging from retail and healthcare to manufacturing (see CBI Insights, Goldman Sachs).

Developing economies, and overall inclusion: There are a number of reasons to be optimistic about the transformative potential of AI in the Global South as the Economist (see Economist) recently highlighted (see also World Bank, IFC). This is also reflected in the generally optimistic views on AI in the Global South (see Google/Ipsos). The advent of smartphones and global communications links helped emerging markets enter the global economy. Similarly, with the right development strategies and digital infrastructure, AI may help empower billions of people. It will in fact be the first major technology to be launched at a time when more internet and mobile phone users live in low and lower-middle income countries than rich ones (see ITU Global Connectivity Report).

AI also presents a unique opportunity to harness the world's knowledge and information, especially in countries where traditional educational resources are not easily accessible. For instance, AI-powered translation is not only helping break down language barriers, but also fosters integration of hundreds of millions of people into the global economy (see Caswell). Already we have seen AI-enabled language translation go from 44 in 2015 to 133 in 2023 and with research efforts now aiming at 1000 languages (see Dean), thereby significantly expanding the inclusion possibilities within countries and across countries and regions. And Google's Project Relate leverages ML to help individuals with non-standard speech to communicate more easily through personalized speech recognition (see Tobin and Tomanek). This is of particular interest given AI's potential to harness knowledge and make it available to people everywhere, and in increasingly skill- and linguistically-accessible ways. With the right investments in digital infrastructure (and electricity in some cases), AI not only offers the potential to deploy accessible products and services, but also an opportunity for many to leverage information, in ways that could generate large economic gains and contribute to economic and societal development (see Economist). As many have highlighted, while there are potential opportunities for the Global South, major gaps will need to be addressed (see United Nations, Brookings).

The inclusion opportunities from AI are not only for developing countries. As became apparent especially during COVID, there are persistent digital divides even in developed countries for some regions and communities, often on the basis of gender, age, and proximity to population centers (see Gallardo, Pew). Moreover, as we explain below, because access depends on investments spanning a multitude of dimensions (e.g. digital literacy, data availability, and reliable electricity) resolving such divides presents a multidimensional challenge. Given AI's potential to assist people, entrepreneurs, businesses and other organizations, and enable more use, economic participation and access to opportunities for people and communities, it will also be critically important that such digital divides within countries are tackled.

### (c) Accelerating scientific breakthroughs

Across a wide range of scientific disciplines, AI can be leveraged to help make discoveries, accelerate scientific breakthroughs, and advance the next generation of scientific methods that can lead to prosperity and well-being for people everywhere (see Department of Energy, Wang et al, Nature and The Royal Society). These methods are not limited to a single aspect of scientific research. Rather, the advance of high-performance computing, large data sets, powerful machine learning and deep learning methods is already transforming scientific discovery across a multitude of scientific fields and applications. Below are some examples from a few scientific fields:

Life sciences: Significant advances in AI can help propel a biotechnology revolution and unlock medical breakthroughs that will pave the way for a more sustainable future for humans. This can be achieved by supporting the rapid development of vaccines and pharmaceutical drugs, engineering new living systems, and enabling the development of new gene therapies.

Google DeepMind's AlphaFold (see Jumper et al) for instance demonstrates the role that AI is increasingly playing in accelerating and disseminating research in nearly every field of biology. AlphaFold predicts a protein's 3D structure from its amino acid sequence, helping solve scientific challenges ranging from advancing new treatments for disease to breaking down single-use plastics. AlphaFold thus solved the 50-year grand challenge of predicting protein structures and led to the prediction of 200 million proteins, saving hundreds of millions of years of research time. Since 2020 when the AlphaFold's protein predictions were made available in an open-access database they have been accessed by more than 1.6 million researchers in more than 190 countries, many of them tackling neglected diseases. And the AlphaFold methods paper has received enough citations to place it in the top 100 most cited papers of the last decade.

AlphaMissence, an AI tool, has also been successfully applied to help uncover the root causes of disease by determining which human genetic mutations could give rise to illness. With millions of possible mutations and limited experimental data, AI is a powerful tool for approaching human genetics. In fact, AlphaMissence is already helping researchers address such challenges by classifying the effects of 71 million mutations (see Cheng et al) and Pangenome, the first draft reference pangenome, reached a breakthrough in our understanding and representation of human genomics through the application of AI (see Liao et al).

Material science: The ability to design materials and chemical compounds has been essential to the rapid advancement of society's technology and physical infrastructure. However, due to the vast number of combinations of elements and a limited understanding of the complex relationships between material properties, the discovery of new materials has often been slow and costly. AI can dramatically accelerate exploration in materials science and expand the universe of potential materials. For instance, GNoME, a deep learning AI tool recently enabled the discovery of 2.2 million new crystals, equivalent to nearly 800 years' worth of knowledge, that could power future technologies (see Merchant et al).

Climate, earth and environmental science: Climate science is vital to tackling the challenges posed by climate change. Recent events, such as intense droughts, devastating storms, and frequent wildfires, have demonstrated our vulnerability to climate change and highlighted the importance of developing climate change strategies. Over the past decade, the emergence of several hundred petabytes of real-time earth data and high-powered computing resources have presented an unprecedented opportunity for AI to not only improve weather modeling – important for social and economic reasons (see Lam et al) – and climate modeling, but also help develop monitoring, mitigation and adaptation strategies as well as energy consumption and efficiency best practices.

Several AI applications are already being used to advance such goals in three key scientific areas. First, AI can improve climate modeling and monitoring by helping predict weather patterns, climate trends, and extreme events (see Wong, Lam et al). For instance, ClimateTRACE: a coalition of nonprofits is using computer vision on satellite imagery to develop an interactive global map of carbon emissions (see Gordon et al). Second, AI can help develop mitigation strategies to reduce greenhouse gas emissions and adaptation strategies to build resilience to the changing climate. Tools like Flood Hub, which uses advanced AI and geospatial analysis to provide real-time flooding forecasts in over 80 countries, and wildfire boundary mapping, deployed in Google Maps and covering more than 550M people, helps users respond to flooding in real-time (Nevo et al). Recent research shows the potential to use AI-based predictions to reduce contrails, the condensation trails that are often seen behind airplanes and account for approximately a third of aviation's global warming impact by as much as 54% (see Geraedts et al). Third, AI can be leveraged to optimize energy production and efficiency by integrating renewable energy sources and making energy generation more efficient. Machine learning can be used to increase the value of wind energy production by predicting wind power output and helping set power generation commitments (see Elkin and Witherspoon). The energy consumption associated with training AI systems has also started to come into focus in recent research (see Luccioni et al). Therefore work to develop more computationally-efficient models will be critical to reduce energy use. AI may also help address the energy footprint of its own models and training processes. Research around reducing ML training energy consumption suggests that if the entire ML field adapts best practices, total carbon emissions from training can decline by 2030 (see Patterson et al).

In addition to modeling and helping mitigate climate change, AI is also used to advance scientific solutions addressing biodiversity loss, deforestation, habitat modification, and species distribution shifts. For instance, in Australia's Great Barrier Reef, machine learning is being used to analyze underwater images of starfish to help combat the emergence of predatory species (see Chung). And because of AI's powerful capacity to predict patterns and analyze satellite images, it is leveraged to monitor deforestation (see Maes) and study species biodiversity and track at-risk species (see WWF).

**(d) Addressing pressing societal challenges**

Beyond AI's potential to advance affirmative benefits, such as productivity, economic growth and opportunity, and scientific discovery, it may also help us address the pressing societal challenges and central threats to our collective prosperity and social wellbeing (see Chui et al). A common framework for understanding pressing global concerns is the U.N.'s Sustainable Development Goals, a collection of objectives agreed upon by 193 countries designed to serve as a "shared blueprint for peace and prosperity for people and the planet, now and into the future." The framework includes 17 goals ranging from ending poverty and hunger to ensuring good health, peace, and justice (see United Nations). Indeed the UN and others have highlighted the potential for AI to help contribute to addressing the SDGs (see United Nations, Department of State). However, the world has reached the halfway mark towards the target date for achieving the SDGs, and more than 80% of its targets are off track. While AI will not be a silver bullet to achieve the goals, fully harnessing its potential to help accelerate progress alongside other initiatives will be important.

For instance, recent advancements in AI have enabled it to play an increasing role in healthcare and medicine. Early-stage adoptions suggest that AI-driven advances can play a significant role in drug discovery, disease detection, and personalized healthcare. At a time when progress towards universal health coverage is facing setbacks and 4.5 billion people (about half the world's population) lack access to essential health services (see World Health Organization), AI can make public health more effective and accessible. It can be used to expand access to basic healthcare services in developing countries through portable applications such as prenatal ultrasounds (see Gomez et al) or automated tuberculosis screening in places where other systems and infrastructures may be lacking (see RSNA).

Advances in AI also present an opportunity to help tackle challenges in education around the world, especially in places where traditional educational resources and infrastructure and sufficient capacity for personalization are lacking. It holds the potential to help foster personalized education, broaden accessibility to educational resources, and confront disparities in education. For instance, ML has been used to not only track learning progress of at-risk students, but also to predict when a student tries and repeatedly fails at an educational task (see Mu et al). And as college enrollment declines (see Meyer), AI has shown the ability to increase graduation rates at a liberal arts college by 32% through tracking students' grades and course selection (see Barron).

And as noted in the prior section, AI is already helping monitor and mitigate climate change, and build a more sustainable food system. In Ghana, AI applications are being used to detect locust outbreaks and identify breeding grounds by aggregating various data sources on rainfall and temperature (see FAO). And in the U.S., AI is being used to predict food security by processing climate data and health indicators (see Feeding America).

To sum up, while we are optimistic about AI's potential to promote the mechanisms of economic progress described above, we also want to highlight the possibility of AI benefits that we simply cannot envision today. AI may even usher in a moment in history when a new tool or way of thinking presents a categorical departure from what existed before. And as we detail in the next chapter, to fully enable the possibility of such a moment, extensive investments in associated process and infrastructure innovations will be needed.

2.  WHY THE ECONOMIC BENEFITS ARE NOT GUARANTEED

As optimistic as we are about AI's incredible economic potential, we also recognize that realizing it is not automatic or guaranteed. The history of innovation indicates that capitalizing on technology's potential for the economy and inclusive prosperity requires effectively tackling several key obstacles and challenges. The economic success of AI may likewise depend on society's ability to take collective action to address several key challenges. We describe below four challenges that must be addressed collectively – by the private sector, public sector, policymakers, those involved in workforce readiness, and other stakeholders – in order to realize AI's economic potential: (a) Sustained investments in R&D, co-innovations and organizational and process changes, (b) Comprehensive and widely-accessible digital and AI-enabling infrastructure, (c) Workforce readiness, transitions and adaptations, and (d) Adoption everywhere and shared benefits.

**(a) Sustained investments in research and development, co-inventions, and organizational and process changes**

A scientific or technical breakthrough — the discovery of an idea — is only the first step in the long process of innovation. As one of Google's founders, Sergey Brin, was fond of saying, "ideas are easy; execution is hard." The successful translation of a scientific discovery from the lab to real-world use requires the effective application of new knowledge, the creation of viable products, the capacity to scale solutions, and a path to reach users who can capitalize the technology.

History is replete with tremendous scientific discoveries that made new products feasible, but did not lead to successful product development (see [Ridley](#)). The Concorde made its maiden flight over half a century ago, yet today not a single commercial flight travels at supersonic speeds. Fusion energy was proven theoretically feasible nearly four decades ago but has not resulted in commercial use. And while humans walked on the moon over 50 years ago, reaching Mars remains no more than a distant aspiration. Many scientific advances such as nanotechnology and quantum computing have thus far generated tremendous scientific discoveries but have not yet advanced sufficiently or led to significant market applications.

AI is subject to the same challenges. Without a continual commitment to invest in research and development including R&D partnerships (between the private sector, academia and public sector in both basic and applied research), focusing product development on useful applications that solve real world problems, create opportunities or enhance productivity, and on products that are usable by many people or organizations, AI scientific and technical breakthroughs may not lead to wide economic and societal impact.

Complementary co-investments: In addition to core investments in product development, new technologies also require complementary investments in new processes, intangible assets, and business models (see Brynjolfsson et al). For instance, the full advantage of gas-powered tractors wasn't realized until farmers developed new crops, new planting and harvesting techniques, and new approaches to processing crops and textiles. Each development required both rethinking of existing approaches and investments to make them successful. Similarly, the "Solow Paradox," coined by Robert Solow in the late 1980s stated that "one can see the computer age everywhere but in the productivity statistics" is often explained by the long time that it took associated process, organizational and infrastructure innovations across large sectors of the economy to deliver the economic impact of information technology (see Mckinsey Global Institute study chaired by Robert Solow). And the benefits of the British Industrial Revolution were delayed by an "Engels' Pause," a half-century-long period of industrial innovation paired with wage stagnation that was partially caused by delays in capital accumulation, co-innovations, and labor market adjustments (see Allen).

The need for such co-innovations and changes, which are often risky and contingent on confidence in the core scientific and technical advances, or coordination of innovations on multiple inter-dependent fronts, can play a critical role in the time that it may take for the full deployment of powerful new technologies. For AI to expand the footprint of digital technologies, businesses in different industries would need to invest in co-innovations to harness AI's full potential. This may require large intangible investments in business models, operational processes, production and other capital assets, employee training, and/or integration of data systems.

Process and organizational changes: The value proposition of a new technology (and hence the likelihood that a user would invest in its implementation) depends on the scope of the solutions that it offers. AI for instance may be leveraged as either a point solution (e.g., enhancing sales techniques, improving employee training, or automating other individual functions) or as a system-wide innovation (i.e., incorporating a wide range of different corporate functions). While both may be beneficial, unlocking AI's full value proposition may require process and organizational changes such as data systems integration or new decision-making workflows.

Electricity provides an example of such dynamics. Before electricity could enable dramatic manufacturing productivity increases, manufacturers had to redesign their factory floors, shifting from a design based on a single source of power (like a water mill) to a new model where power was available at any point in an assembly line. Taking full advantage of the new technology required revamping decision-making processes and organizations (see Agrawal et al).

Similarly, organization and process alignment was also a differentiating factor in the adoption of digital technologies. In fact, leading tech companies did not emerge solely because of their willingness to invest in IT or hire employees with digital acumen, but also due to their ability to integrate IT systems, make organizational innovations and changes, and align new processes with key IT investments, as was also the case in "big-box" retail and wholesale sectors in the 1990s (see Farrell et al).

Healthcare as well demonstrates that similar process and organizational changes may be required to unlock the full value proposition of AI. Hospitals are increasingly relying on AI to improve specific functions such as medical imaging or management of medical records. While these point-solution applications can certainly improve the performance of an individual hospital function, they do not necessarily tap into the full potential of AI as a system-wide innovation. The latter may require a hospital to integrate data systems, reorganize decision making, and re-envision the modularity of its different medical functions. Hospitals that commit to addressing such challenges stand to enhance performance in a number of interdependent fields thereby improving clinical decision making, patient flow, and ultimately overall outcomes (see Agrawal et al).

### (b) Comprehensive and widely-accessible digital and AI-enabling infrastructure

In the past decade the world has experienced a technological transformation that is unprecedented in its depth and speed. Within a single decade internet access was expanded to 2.3 billion people (see ITU Connectivity Report). As a result, the developing world, which accounted for only a third of the digital population two decades ago, today stands at about 72%. This transformation presents a tremendous opportunity, as AI will be the first major technology to be launched at a time when more internet and more mobile phone users live in low and lower-middle income countries than rich ones.

However a significant digital divide persists, as nearly three billion people are either offline or or do not have sufficient access to take advantage of the opportunities today and many users have access to only basic connectivity (see United Nations, ITU). And while more than 90% of people in high-income countries used the internet in 2022, only one in four in low-income countries have access (see ITU). After decades of private and public investments in infrastructure, internet access remains persistently challenging with large shares (e.g. Africa 60%, Asia Pacific 36%) of the population having either unreliable, unaffordable or no access at all to the internet.

Failing to invest in wider digital infrastructure implies that AI's tremendous benefits would be shared with only a subset of the world's population. PwC for instance estimates that nearly 70% of economic value generated by AI will accrue to just two countries: the U.S. and China (see PwC). This imbalance raises concerns about growing global disparities in the economic benefits of AI, and without deliberate efforts to address the imbalance, it will likely worsen over time. This is particularly concerning given AI's ability to advance public health, sustainability, and economic opportunities that may be relatively more impactful in lower-income markets.

Collaborations encompassing local governments, private sector actors, telecom carriers, and/or digital firms have often worked to expand digital infrastructure. For example, with Google's Equiano, a subsea cable from Portugal to South Africa (see WIOCC), Humboldt, a subsea cable connecting Chile to Australia (see Department of State) and Seacom, Africa's first broadband submarine cable system (see Moyo), privately funded infrastructure is managed through partnerships between local companies and global organizations. And investments in new broadband technologies, such as Taara, which is developing invisible beams of light to transmit information at high speeds, may help improve broadband cost-effectiveness and expand the infrastructure boundaries (see Lee and Frandino).

Incomplete digital infrastructure is however a multifaceted challenge. Ensuring access involves more than just having fixed or mobile connectivity available. It also includes other AI-enabling infrastructure such as service continuity, device affordability, data availability, compute power, and digital skill proficiency. For instance, even in areas where the infrastructure can support broadband connectivity, data imbalances may undermine AI proliferation. Despite the tremendous growth in data and its cross border trade, a data imbalance persists as many countries lack adequate access to both private and public data (see World Bank, Sudan et al). And while broadband infrastructure requires costly investment, the resources and collaborations required to reach universal meaningful connectivity are even greater, particularly in areas that are less densely populated, require greater infrastructure investments, or offer lower private returns (see ITU). For these reasons, and as we discussed above, disparities in internet connectivity are not limited to the Global South, as in many advanced and developing countries digital divides persist within countries as well.

### (c) Workforce readiness, transitions and adaptations

The potential for AI to benefit the economy and workers may depend on the resilience of the labor market, workforce readiness, the capacity for workers to work alongside assistive or augmenting AI, avoidance or mitigation of potential adverse wage effects, and effective transitions and adaptations for any potential displacement. AI's benefits may be limited if it causes job losses and workers can't find new jobs, the workforce lacks the skills needed for an AI transition, or AI is implemented in a way that replaces workers instead of augmenting their capabilities. We discuss these concerns below.

Impact on individual employees: We discuss in the next chapter that based on most research, it is unlikely that AI will result in a widespread loss of jobs or mass unemployment. As we note, most studies conclude that more jobs will be gained than lost in the next decade or more (e.g., McKinsey Global Institute, Acemoglu et al, D'Andrea, Tyson, and Zysman). Yet, even if AI creates enough net new jobs to offset those that it may displace, some workers may lose their jobs during the transition, need to adapt to AI-assisted work, or need to find new jobs altogether. And while AI may eventually create better jobs, there is no guarantee that displaced workers would be reemployed, have the right skills, or that better jobs would emerge in time.

Such challenges have been seen before. During the 20th century, as technology and automation contributed to dramatic gains in productivity, overall employment as a share of the population actually increased. At the same time, the United States workforce endured a profound shift in sectoral employment as the agricultural share of total employment declined from 60% in 1850 to less than 5% by 1970, and employment in manufacturing fell from 26% in 1960 to less than 10% today (see Lund and Manyika). Although overall employment did not decline, there were profound and often challenging transitions of workers from declining occupations and sectors to those that were growing. Hence the imperative to focus on effective workforce readiness and mechanisms to ease the likely transitions.

Workforce transitions: Concerns about potential displacement of employees may not only harm individuals, it can also impede the broader transition to realizing the beneficial potential from AI. How do we collectively grow new jobs for displaced employees, ease worker transitions, develop education programs that provide adequate training for needed skills in new occupations or for working alongside rapidly-evolving AI?

Developing AI proficiency and avoiding skill atrophy may be particularly important during the transition. Even in ordinary times, skilling programs face a variety of challenges, including prohibitive participation costs, scalability difficulties, and the need to keep up with rapid technological advancements and market conditions. Technological disruption and uncertainty about the nature of future jobs can exacerbate these challenges.

These issues have been observed before, in prior attempts to address globalization-induced labor displacement (see World Trade Organization). While globalization produced significant consumer and economic benefits, many trade adjustment assistance programs for workers suffered from inherent limitations and proved insufficient. In particular, enrollment in such programs often required a temporary income sacrifice, as participants invest in skills development with the expectation of future earnings growth. Empirical evidence however suggests that such prospects are often uncertain and ineffective at raising long-term compensation. To be effective, AI-transition assistance programs would need to carefully address such limitations.

Human-complementary AI: Another challenge relates to whether AI will encourage substitution of human capital because of its unique features. Technology can serve as both a substitute (replacing human work) and as a complement (assisting human work and making it more valuable). But when technology development is narrowly focused on matching or replicating human capabilities (as opposed to complementing human capabilities) it can lead to substitution. For instance when a new technology is used to reduce costs by merely replicating a human task or developed to benchmark and mimic an individual human capability, it may lead to displacement. And it has been shown that even when technology is developed to complement rather than substitute, some users may employ it as a substitute in an effort to reduce costs This tension has been referred to by Erik Brynjollffson as the Turing Trap, which, in the case of AI, can be triggered by an excessive focus on creating human-like AI, narrow use that displaces workers as a result of cost-cutting pressures, or policy and other incentives that encourage substitution (see [Brynjolfsson](#)).

The challenge is whether AI can be leveraged in a more comprehensive way that encourages complementing human capacity. For instance, complementing rather than substituting human work creates new capabilities, products and services, potentially generating far more value to both workers and businesses (see [Bughin and Manyika](#)).

Some have argued that whether AI complements or substitutes work, is a matter of choices and policies (see [Acemoglu and Restrepo](#)). As we discuss in the next chapter, tackling the Turing Trap will require wide stakeholders' involvement (i.e., by the developers of AI technologies, and the companies, organizations and employees that use them), as well as policies that can shape how AI is developed, deployed and used.

### (d) Adoption everywhere and shared benefits

Uneven adoption of AI could limit the potential for broad productivity gains and stall economic growth. To ensure economy-wide productivity gains, a sufficiently large number of sectors, especially large sectors, and a broad base of businesses would need to adopt productivity-enhancing AI applications. Otherwise, if limited access, poor infrastructure, insufficient data, weak business relevance or other barriers (such as a reluctance to co-investment in innovations, organizations and processes) confine AI adoption to a subset of frontier firms or limited number of small sectors, AI will not generate a significant economy-wide impact (see [Manyika and Spence](#), [Bank of England](#)). In the U.S. for instance, as employment in the digital economy (defined as workers for companies focused on digital services and applications) accounts for only 5% of total employment, a failure to leverage AI to large services, manufacturing or traditional industries will not generate meaningful productivity gains (see [Bureau of Economic Analysis](#)). This is one of the key lessons from previous eras of technological advancement: Until they impact and transform large sectors of the economy in productivity-enhancing ways, their impact on overall productivity growth and economic growth is limited (see [Robert Solow chaired study](#)).

The history of digitalization suggests that it has not been uniformly applied across different sectors of the economy (see [Manyika et al](#)). For example, in the U.S., while some sectors (e.g., media, professional services, and finance) were quick to digitize, other sectors (e.g., healthcare, construction, and agriculture) have continually fallen behind. The uneven distribution of innovation was not without consequence. Over the last few decades, it gave rise to lingering innovation bottlenecks, which not only served as a primary cause of an economy-wide productivity slowdown, but also contributed to slower economic growth and hindered the realization of the benefits and economic impact from digitization (see Acemoglu et al).

And similar gaps exist globally at the company-level as well - both within sectors and across sectors. While some superstar firms (which are not always large firms) have made successful investments in innovation, many others — particularly small and medium-sized businesses, are reluctant to invest, face significant barriers to adoption, or lack the willingness to make the needed process and organizational changes to take advantage of the technological advancements (see Autor et al, Zolas et al, Manyika et al, Bank of England). Amongst the world's largest cities, the divide between "digitally smart" cities that emerge as technology hubs and counterparts that lack digital infrastructure is expanding. If adoption obstacles persist in these sectors, businesses and cities, uneven adoption may once again become a fundamental threat to AI's ability to deliver shared economic prosperity.

Uneven AI adoption may also lead to disparities in economic impact and benefits. Because technology serves as an extraordinarily powerful mechanism to drive economic prosperity, it tends to have distributional consequences. Income distribution is of course important, not only due to concerns over equity or fairness, but also because uneven distribution is often the product of limited opportunities, economic exclusion, and/or mobility barriers. How such prosperity is shared, and whether it advances new opportunities, will play a role in society's willingness to embrace AI.

While technology and other market economy innovations have driven tremendous increases in prosperity over the last century, the benefits have not been evenly distributed (see Manyika et al), reflecting wider factors such as public policies, social norms, and socio-economic inclusion and mobility. This is also consistent with the significant differences in income inequality we see across different countries (see PIIE). A comparison of the U.S. and Europe for instance indicates that the two economies have very different levels of inequality despite similar levels of technological adoption, suggesting that technology may be a contributor to income inequality, but not the main cause. Government interventions and public policies often play a significant role in income distribution (see D'Andrea, Tyson, and Zysman, Benedikt Frey).

But the way we adopt technology can also have distributional consequences in light of mobility barriers and existing market institutions. Skill-biased technological progress often led to disproportionate benefits flowing to more educated and more skilled workers (see Autor). In fact, the impact of AI on white-collar, blue-collar, and "new-collar" jobs remains uncertain — just as industrialization both complemented and substituted for some blue-collar jobs, AI may have similar effects on white-collar employment.

In summary, the history of innovation demonstrates that the most impactful technologies often face prolonged challenges before they become ubiquitous. While Edison demonstrated the electric light bulb in 1879, it would take another four decades before most homes became electrified and were able to install light bulbs. While the Wright brothers pioneered the first manned flight in 1903, it took another half a century for commercial aviation to become widely available. More recently, the insights from digitization suggest that while a technology can seem to be everywhere all at once, until it has been adopted and used broadly in productivity-enhancing and opportunity-expanding ways across large swathes of the economy, overall economic impact and prosperity will not be fully realized. Technological change has of course dramatically accelerated since Edison's light bulb or the Wright brothers historic flight. But they serve as vivid reminders that transformative technologies often face persistent challenges that can take time and collective action to overcome.

3.  WHAT HISTORY SUGGESTS ABOUT TECHNOLOGY AND EMPLOYMENT

We discussed above the economic benefits and challenges facing AI, and now turn to consider the likely impact of AI on employment. We consider two questions that may help address this issue: (1) Are digital technology and automation more often a substitute or complement to human labor, and (2) Have digital technology and automation to date resulted in mass job loss or unemployment?

**Technology tends to replace tasks, not jobs**

The automation of work over the last few decades has shown that there is an important distinction between tasks and occupations. Where a task is a specific, well-defined unit of work, an occupation consists of the agglomeration of a wide range of tasks, activities, and responsibilities. The distinction is important because technology tends to replace individual tasks, not entire occupations. (It typically leads to only partial automation of any given occupation.) That's why very few occupations are typically lost even over prolonged periods of time. For instance, according to one study, of the 270 occupations listed in the 1950 U.S. Census, only one, elevator operator, was eliminated due to automation (see Bessen, Kessler).

AI seems to be tracking the pattern of earlier innovations, automating a subset of tasks but typically not entire occupations (see Agrawal et al). For instance, in transportation it frequently automates navigation, but (at least so far) rarely all of driving; and in medicine, it more often automates diagnosis but infrequently care. As we expand the capabilities of AI to replace more cognitive and non-routine tasks, the subset of tasks that can be automated will naturally shift over time. While this should raise concerns, it does not necessarily imply a massive loss of jobs, but rather that human work is changing through task-specific automation.

In the cases of driving, AI-powered maps have actually enabled platforms such Uber and Lyft to dramatically increase the number of drivers for hire. While in healthcare, the growth in AI-enabled medical imaging has actually *increased* the demand for radiologists. In fact, with the increased integration of AI as an advanced tool for image analysis and diagnosis (see Agarwal et al), AI has helped enhance tumor identification, streamline workflows, and improved treatment decisions, but it has not replaced human radiologists. It actually led to a surge in demand for radiologists in the past few years, even causing significant shortages in their workforce (see Montecalvo). Hence technology in these examples not only replaced individual tasks, but it actually led to greater employment in impacted occupations.

**Technology often complements human work, but not always**

As we noted above, whether technology serves as a substitute (replacing human work) or as a complement (making human work more valuable) is an important determination of whether jobs will be created or lost.

Enhancing the value proposition of labor: While technological progress is often motivated by the desire to replace ordinary human tasks (e.g., plowing a field or working on a factory line), technology can lead to both substitution and complementarity with labor. Focusing solely on substitution overlooks a central economic mechanism by which automation may actually enhance the value of human work (see Autor). When it replaces tasks that are mundane, reduces drudgery, or improves workers' autonomy, technology can improve labor's value proposition even if it replaces certain tasks. It empowers humans to focus on human-unique traits, and augments the use of critical thinking and creativity. That not only increases the demand for labor, but also frees people to pursue more meaningful and rewarding work.

Combining humans and machines: Technology on a stand-alone basis is often at a disadvantage relative to the combination of technology plus human labor. In fact, we often see that even when machines outperform human competencies in specific tasks, they often do not outperform the combination of person plus machine in complex real-life environments (see Cao et al, Autor et al). Moreover, the combination of human and machine can also generate benefits through comparative advantages. It can specifically enhance the productivity of both labor and machine by enabling them to focus on their relative advantages (i.e., the respective tasks that they are uniquely good at). For AI that may be pattern recognition and prediction. For humans that may be critical thinking and decision making. Efficiently allocating tasks between humans and machines frequently improves labor productivity and makes human work more valuable.

Creating new jobs: Technology can also enhance the value proposition of human work by shifting employees to new, more productive occupations. U.S. occupational changes in the 20th century provide us a remarkable account of this dynamic. Since the 1940s, new technologies have increasingly automated a large number of human-performed tasks. As a result, 60% of the U.S. workforce today is in occupations that did not exist eighty years ago (see [Autor et al](#)). It is also important to note that technology-fueled job creation is not limited to the emergence of new job categories that didn't exist before, but is also generated by growth in demand for existing categories of jobs, as several studies have shown (see [Manyika et al](#)).

Just to illustrate this point, if we traveled back in time to the peak of the dot-com era and grabbed a copy of contemporary labor statistics, they would look shockingly incomplete. They would specifically exclude new roles (like web developers, UX designers, and social media managers) and entirely new sectors (like e-commerce, and cloud computing) that were created by the subsequent growth of the internet. This illustrates the power of technology to create jobs and introduce new occupations even within a relatively short period of time.

But, it is worth noting that employment in some occupations have certainly declined over time even as output increased in some cases. For instance, as we noted above, agricultural share of total employment declined from 60% in 1850 to less than 5% by 1970, and employment in manufacturing fell from 26% in 1960 to less than 10% today (see [Lund et al](#)), but in both cases employment shifted to other growing services sectors.

**Technology has not led to massive job losses or unemployment**

Throughout recent history, technological advancements have been met with fears that they will make human labor obsolete. And yet, the prospect of technology triggering economy-wide unemployment or net loss of jobs runs contrary to the evidence. During the 20th century, the most technologically-innovative century in human history, unemployment did not increase, and the employment-to-population ratio did not drop. In fact, as labor markets endured tremendous change (including the absorption of women into the workforce and a transition from agriculture to services and manufacturing), technology led to a net job creation and persistent rise in employment (see [Autor](#)). And as demand for labor expanded, new occupations grew, the services sector and the global economy expanded and productivity growth helped create a virtuous cycle. These findings led most economists to agree that, as discussed above, automation can certainly disrupt, lead to sectoral changes and exacerbate income inequalities, but it has  not led to mass loss of jobs or unemployment. This view is reflected by a number of studies (see [MIT](#), [Bureau of Labor Statistics](#), [McKinsey Global Institute](#), [Acemoglu et al](#), [Autor](#), [Goolsbee](#) and [Benedikt Frey](#)). It is telling that nearly seven decades ago, when concerns over automation led President Lyndon Johnson to assemble a National Commission on Technology, Automation, and Economic Progress, it too recognized "the basic fact that technology destroys jobs, but not work." This view is further corroborated by a number of case studies (see [National Commission on Technology, Automation, and Economic Progress](#)).

In the early 20th century in the U.S. the introduction of automobiles sparked public concerns regarding job loss. Indeed, between 1910 and 1950, over half a million jobs were displaced by automobiles (see Manyika et al). This included occupations such as wagon and carriage manufacturers, harness and saddle makers, and metal workers. However, automobiles did not solely result in job displacement in traditional transportation segments. They also generated a significant influx of new employment opportunities, including roles in car manufacturing and supply chain (e.g., metal parts manufacturers and warehouse employees), as well as in industries supporting automobile use (e.g., auto dealerships, auto repair shops, and gas stations) and subsequently in new economic sectors (e.g., long-haul trucking and interstate highway systems). All told, automobiles are estimated to have generated a staggering 7.5 million jobs, which are not only roughly tenfold the amount of jobs that they displaced, but also an estimated 10% of employment at that time. While the concerns over displacement were evident and genuine at the time, it was probably difficult to anticipate the tremendous job creation that automobiles would eventually stimulate in subsequent years.

Similarly, the growth of personal computers in the 1970s had a significant impact on U.S. employment (see Manyika et al). Digital editing, for instance, made it easier for authors to type and edit documents directly without the need for people or machines that are specialized in typewriting. As a result employment for typists, typewriter manufacturers and secretaries fell significantly following the advent of personal computers, ultimately leading to a loss of 3.5 million jobs. But the growth of personal computers actually created far more jobs than it destroyed. At least 19 million jobs were actually created between 1970 and 2015 in a wide range of new and existing occupations including not only computer manufacturing, but also IT administrators, software engineers and call center employees.

And, while the advent of ATMs in the 1970s was feared to render bank tellers obsolete, it actually led to an increase in the aggregate number of bank branches and demand for bank tellers, in the subsequent four decades. This was largely because though the number of bank tellers per branch declined, overall demand for branch banking grew and drove demand for bank tellers and related occupations. Moreover the role of the back teller also evolved beyond simply dispensing cash, to focus on a wider range of customer service functions (see IMF). The trend of growth in aggregate demand for bank tellers eventually declined after four decades, not because of the advent of the ATM, but because of another shift, the growth of online banking and lower usage of cash. And as e-commerce surfaced in the early 2000s, many feared that it would trigger mass losses of jobs (see Mandel). Yet, not only did e-commerce generate a large number of jobs (in some cases converting personal tasks like running errands to paid labor like shipping and delivery), it also contributed an increase in pay relative to traditional brick and mortar.

To sum, the history of technology provides much to be optimistic about. Over the course of the last century, technology has not increased unemployment or led to widespread losses of jobs. This finding is consistent with the notion that technology tends to replace tasks, not jobs and can create new work streams that complement and enhance human work. This seems particularly relevant for AI which, based on early case studies, has a tremendous potential to augment human capital, enhance productivity and improve the value proposition of work.

However, while there are many reasons to be optimistic about AI's role as a catalyst for better jobs and what The Economist recently described as "a golden age for workers" (see The Economist), given technology's current interplay with jobs and the state of the labor markets in most advanced economies, we also recognize that the challenges discussed earlier – of workforce readiness, transitions, possibility of displacement – are significant and important to tackle. Moreover, AI may introduce new challenges, such as the possibility of change occurring faster than in previous eras. Next, we consider some key foundational steps to tackle the potential challenges.

4. WHAT KEY ELEMENTS OF AN AI POLICY AGENDA ARE NEEDED

Scientific and technical breakthroughs in and of themselves do not guarantee progress or economic impact and prosperity for all. In fact, the ability to capitalize on AI's economic promise, manage the transition to an AI-empowered economy, and help prepare for the future, depends on developing balanced policies that address the significant challenges that we detailed above.

As Google notes in a recent white paper, to capitalize on AI's economic promise, a balanced policy agenda must include three central pillars: Responsibility, security, and economic opportunity (see Walker). Responsibility is essential for its own sake to ensure safety, trust and confidence in the widespread adoption of AI technologies. Security is critical to prevent exploitation and misuse by malicious actors. And maximizing AI's economic dividends and distributing them broadly requires a focus on economic opportunity – which has been the subject of this paper,

The historical record presented above underscores the need for a comprehensive and collective economic opportunity agenda, which should include three central elements: (a) Investing in AI infrastructure and an innovation ecosystem; (b) Focusing on workforce readiness, building an AI-assisted workforce, advancing AI as a human-complementary technology and easing the challenges of workforce transitions, and (c) Promoting widespread adoption and addressing barriers to organizational and process changes to capitalize on AI potential. While more may be needed, given the novelty of AI, if we take lessons from history and what we can already see seriously these three elements are critical and foundational as we move forward.

Moreover, to maximize the upside potential and mitigate the downside risk requires focus and investments across the private and public sectors, including from developers of AI, the deployers of AI, the users of AI (SMBs, large corporations, public sector and other organizations), the providers of digital and other infrastructure, those involved in workforce education and training, as well as from policymakers and civil society.

### a. Investing in AI infrastructure and innovation ecosystem

The health of a country's innovation ecosystem can play a critical role in determining whether AI realizes its full economic potential. Such an ecosystem includes the legal and institutional environment needed to promote investments, facilitate scientific collaboration and enable product development. AI presents unique challenges that require coordination across different constituents (including academics, policymakers, technology firms, startups, and labor). A number of the policy initiatives highlighted in Google's AI opportunity agenda – enabling legal frameworks, STEM investments, technology transfer, and public-private-academic partnerships – are needed to strengthen an AI innovation ecosystem.

We also described above the importance of AI in advancing universal accessibility and shared prosperity. To ensure that AI advances such goals, it is essential that it does not become the exclusive domain of a select few: The digital infrastructure that supports AI must be available to all users and businesses, across all sectors and economies. It will be important to enable access to compute, tools and other infrastructure in order to have an open, rich and vibrant ecosystem of many AI model developers and AI applications developers to develop useful applications and solutions for a wide variety of needs for a wide variety of use-cases and business functions in multiple industries and regions (see Ichhpurani). As part of fostering such a broad and inclusive innovation ecosystem, it will be important to establish responsible approaches to open-sourcing of AI models (see Klyman). Moreover, meaningful internet access is a fundamental infrastructure needed to reach users, especially in the Global South where many still do not have access to these critical enablers. To accomplish wide accessibility will require policies, initiatives and public-private partnerships that encourage infrastructure investments.

### b. Focusing on workforce readiness, building an AI-assisted workforce, promoting labor complementarity, and easing the workforce transitions

As noted above, the history of technology suggests that it is unlikely that AI will lead to mass unemployment or net loss of jobs, especially if research and advanced applications that promote labor augmentation are prioritized. However, AI will inevitably disrupt the workforce and lead to sectoral and occupational shifts. To ensure that worker displacement is temporary, and that individuals who may need to switch jobs have practical alternatives, an AI-assisted workforce and a workforce transition agenda to develop relevant human capital will be needed.

AI presents one of the central challenges of the 21st century, and the workforce's ability to adapt to an AI-assisted economy is certainly a key component of such a challenge. The transition to an AI-assisted economy however will not happen solely on the basis of goodwill or market dynamics. It would require addressing major workforce readiness, occupational and sectoral adjustments. And while such transition may lead to productivity gains and accelerated growth, it would also require a deliberate and collective strategy. Without such a strategy, humanity not only risks developing AI that is inconsistent with human-complementary growth, but also hindering the capitalization of AI's tremendous economic potential (see Acemoglu and Johnson).

Developing a workforce transition strategy can be challenging in the ordinary course of business, let alone when disruptive technology is changing the future of work and shifting economic sectors. A number of the policy initiatives highlighted around an AI Corps in the AI policy agenda (such as public-private partnerships to scale career certification programs, ongoing updates of skilling programs to prevent skill atrophy, or advancement of apprenticeship programs) set the directions for such strategy and may support a successful workforce transition.

Another workforce transition initiative relates to fostering labor complementarity: When AI tools are used merely to replace human tasks, rather than assist or complement human tasks, this can have adverse effects on jobs. As has been highlighted by Brynjolffson and others, this can be fostered by focusing on how the technology is developed, how it is applied and used, and also what incentives are put in place to encourage complementarity rather than substitution (see Acemoglu et al). Furthermore, to help encourage complementary AI, governments can invest in both (a) applied research that promotes augmentation and human-complementary technologies and (b) policies, incentives and initiatives that encourage businesses to leverage AI in a way that adds value and enhances human capital, including fostering investments that pairs AI tools with human expertise in strong prospect sectors, and/or promoting partnerships consisting of developers, users and labor organizations to coordinate on development, application and use of new technologies in ways that enhance human capital. A recent cross-country comparison of automation indicates that such initiatives can be effective (see Kapetaniou and Pissarides). It found that for a similar set of technologies, countries with a poor innovation ecosystem substitute technology for workers much more frequently than do countries with an advanced innovation ecosystem. These findings suggest that a country's investments in an innovation ecosystem not only accelerate the implementation of new technologies, but also tend to make them more complementary than substitutes for labor.

### c. Promoting widespread diffusion and facilitating process and organizational changes to capitalize on AI's potential

AI technology must be broadly accessible and ubiquitously used. However, across the 21st century, technology's history shows that adoption barriers gave rise to disparities in adoption, as many small businesses and traditional industries used only a fraction of the tools available to their larger, better-positioned peers. AI presents an opportunity to breach this adoption gap by offering new technological capabilities that are more flexible and valuable. But it will be important to ensure that AI applications are rich and diverse, and developed to address opportunities across business functions and sectors of the economy, and that they are accessible and applicable to business use cases that enhance productivity and innovation. This will require a rich and vibrant AI ecosystem of model and application developers, as well as those providing infrastructure and tools, and those enabling deployment and use. Furthermore it will be important that businesses and organizations that use AI make the co-investments and organizational and process changes needed to fully capitalize on AI's potential to enhance productivity and be assistive to their workers.

As discussed above, the drive for wide participation in an AI-assisted economy stems not only from a concern over economic equities, it is also a condition for promoting shared prosperity and advancing economy-wide productivity gains. Hence ensuring that all economic actors including low-income households, small businesses, traditional industries, and manufacturing and services sectors can participate and benefit from AI's potential (in both commercial and non-commercial opportunities) is critical for its ability to generate wide economic benefits. To meet these needs, the challenges detailed above must be addressed, including supporting SMBs in process and organizational changes, developing products that are reliable for wide use, promoting human-complementary AI which does not prioritizes substitution, ensuring that digital infrastructure is complete, and that workers have access to effective, practical vocational programs (see PPI). Enabling widespread diffusion across businesses within and across sectors and accelerating process and organizational changes to capitalize on AI, will require action by all – developers of AI, those providing enabling infrastructure, business and other organizations using AI, all sectors including the public sector, and policy-makers.

A CONCLUSION AND A BEGINNING

AI presents a once-in-a-generation opportunity to advance economic growth and broaden prosperity. It is a foundational new technology that can dramatically change the way that we harness information and knowledge to perform tasks from the everyday tasks to the most creative and ambitious — dimensions that are central for economic progress, for individuals, small and large businesses, sectors and whole economies.

AI exhibits all of the makings of a general-purpose technology, a transformative technology powerful enough to transform and accelerate national and global economic progress across entire sectors and nearly all economies. It is also uniquely positioned to advance specific mechanisms of economic progress including improving labor productivity, expanding economic opportunity, accelerating scientific breakthroughs, and addressing pressing societal challenges. These present a historic opportunity to collectively benefit all of society.

However, AI's potential to transform the economy and drive a shared prosperity that benefits all is not automatic or guaranteed. The history of innovation suggests that capitalizing on AI's potential requires tackling obstacles and challenges, ranging from uneven infrastructure to adoption barriers and organizational changes to workforce readiness and disparities in access to its opportunities and benefits. As discussed in this paper, there are clear lessons from the past and insights from recent studies on AI's potential, however, it's important to note that many of the empirical case studies are still early and limited in scope – therefore more research is needed especially as AI continues to develop, and more is learned from its deployment, use and impact. While we don't possess all the answers, getting AI right – realizing its benefits, while addressing its challenges and risks – will demand that we build on relevant lessons from the past, and at the same time work to address the new emerging questions, challenges and beneficial opportunities, and embrace an affirmative policy agenda that will require collective effort by everyone: developers, deployers, users, infrastructure providers, labor organizations, academics and policymakers.

## Acknowledgements

We thank many of our Google colleagues who have contributed to AI research and innovations we cite in this paper, as well as Google colleagues who contributed to the ideas and insights put forward in this paper, including Blaise Aguera y Arcas, Travis Beals, Nicholas Bramble, Elana Burton, Luke Garske, Lisa Gevelber, Brigitte Hoyer Gosselink, Nicklas Lundblad, Preston McAfee, Anoop Sinha and David Weller. In addition we have benefited from the work of economists outside Google many of whom we have been in dialogues on many of the ideas in this paper, including Daron Acemoglu, Ajay Agrawal, Ben Armstrong, David Autor, Erik Brynjolfsson, Michael Chui, Laura D'Andrea Tyson, Avi Goldfarb, Jan Hatzius, Katya Kilnova, Anton Korinek, Michael Mandel, Andrew McAfee, Heidi Shierholz, Michael Spence, and John Van Reenen as well as many others whose published work we cite. In addition we are grateful to colleagues Kerry McHugh and Michael Pisa, who provided detailed and thoughtful comments on an earlier draft.

The views expressed herein are those of the authors and do not necessarily reflect the views of Google or Alphabet, or those we have been in dialogue with.